\documentclass[a4paper]{article}

\usepackage{INTERSPEECH2021}

\usepackage{enumitem}

\usepackage[utf8]{inputenc}
\usepackage{pgfplots}
\DeclareUnicodeCharacter{2212}{−}
\usepgfplotslibrary{groupplots,dateplot}
\usetikzlibrary{patterns,shapes.arrows}
\pgfplotsset{compat=newest}

\usepackage{url}

\def\R{\mathbb{R}}
\def\Pmiss{P_\text{miss}}
\def\Pfa{P_\text{fa}}

\DeclareMathOperator*{\argmin}{argmin}

\def\R{\mathbb{R}}
\def\Pmiss{P_\text{miss}}
\def\Pfa{P_\text{fa}}

\DeclareMathOperator{\EER}{EER}

\title{Out of a hundred trials, how many errors does your speaker verifier make?}
\name{Niko Br\"ummer$^1$, Luciana Ferrer$^2$, Albert Swart$^1$}
\address{
  $^1$Phonexia, South Africa\\
  $^2$Instituto de Investigaci\'on en Ciencias de la Computaci\'on, CONICET-UBA, Argentina}
\email{niko.brummer@gmail.com,lferrer@dc.uba.ar}

\begin{document}

\maketitle
\begin{abstract}
Out of a hundred trials, how many errors does your speaker verifier make? For the user this is an important, practical question, but researchers and vendors typically sidestep it and supply instead the conditional error-rates that are given by the ROC/DET curve. We posit that the user's question is answered by the Bayes error-rate. We present a tutorial to show how to compute the error-rate that results when making Bayes decisions with calibrated likelihood ratios, supplied by the verifier, and an hypothesis prior, supplied by the user. For perfect calibration, the Bayes error-rate is upper bounded by min(EER,P,1-P), where EER is the equal-error-rate and P, 1-P are the prior probabilities of the competing hypotheses. The EER represents the accuracy of the verifier, while min(P,1-P) represents the hardness of the classification problem. We further show how the Bayes error-rate can be computed also for non-perfect calibration and how to generalize from error-rate to expected cost. We offer some criticism of decisions made by direct score thresholding. Finally, we demonstrate by analyzing error-rates of the recently published DCA-PLDA speaker verifier.        
\end{abstract}

\noindent\textbf{Index Terms}: speaker recognition, calibration, Bayes decisions

\section{Introduction}
\def\Pmiss{P_\text{miss}}
\def\Pfa{P_\text{fa}}
\def\barPmiss{\bar P_\text{miss}}
\def\barPfa{\bar P_\text{fa}}
\def\tildePmiss{\tilde P_\text{miss}}
\def\tildePfa{\tilde P_\text{fa}}
\def\acc{\text{accept}}
\def\rej{\text{reject}}
\def\Cmiss{C_\text{miss}}
\def\Cfa{C_\text{fa}}

This is a position paper and tutorial about the decision stage of speaker verification and how to quantify verification accuracy. Our position is: 
\begin{itemize}
	\item Bayes decisions are preferable to ad-hoc alternatives.
	\item The Bayes error-rate (the accuracy of Bayes decisions) is a good representation of the accuracy of the verifier.
\end{itemize}
A \emph{speaker verification trial} provides two speech samples that may have been spoken by one-and-the-same speaker (hypothesis $H_1$), or by two different speakers ($H_2$). It is required to make an accept/reject decision, where \emph{accept} favours $H_1$ and \emph{reject} favours $H_2$. The speech input of the trial is processed by a speaker verifier that outputs a \emph{score}, $s\in\R$, where by convention, more positive $s$ supports $H_1$, while more negative $s$ supports $H_2$. For some verifiers, this is all that can be said of the scores and such scores are termed \emph{uncalibrated}. Some verifiers output their scores in (log) likelihood-ratio format, termed \emph{calibrated}. Uncalibrated scores can be post-calibrated by adding a calibration stage to the verifier.  

Bayes decisions can be made with either calibrated, or uncalibrated scores as will be explained in this tutorial. The former method provides the user with more flexibility and allows for more informative representation of the verifier accuracy. Decisions \emph{can} also be made by comparing scores to ad-hoc thresholds, obtained e.g.\ by fixing the false-accept rate. While such methods avoid both likelihood-ratio calibration and specification of prior and costs by the user, we offer some strong criticisms.

The tutorial does not discuss how to calibrate scores, but rather how to make decisions given calibrated scores---and how to compute the accuracy of these decisions. We discuss accuracy measures that can be obtained by testing the verifier on a supervised evaluation database. Perhaps the most ubiquitously available accuracy statistic is the equal-error-rate (EER). We highlight a little-known relationship between the EER and the Bayes error-rate: for a well-calibrated verifier, the EER is an approximate upper bound to the Bayes error-rate. More generally, given an ROC/DET curve~\cite{DETPLOT} of the form $\Pmiss(\Pfa)$, the Bayes-error-rate for a well-calibrated verifier can be computed as a function of the prior, $P(H_1)$. Most generally, given $\Pmiss(\theta)$ and $\Pfa(\theta)$ as functions of a score threshold, $\theta$, the actual Bayes error-rate can be computed for verifiers that may or may not be well-calibrated. 

The theory in several sections below is followed by a demonstration of our methods on the recently published DCA-PLDA speaker verifier~\cite{DCA_PLDA}.

\section{Optimal Bayes decisions}
\begin{figure}[tb!]
  \centering
  \includegraphics[trim=0 10 0 10, clip, width=\linewidth]{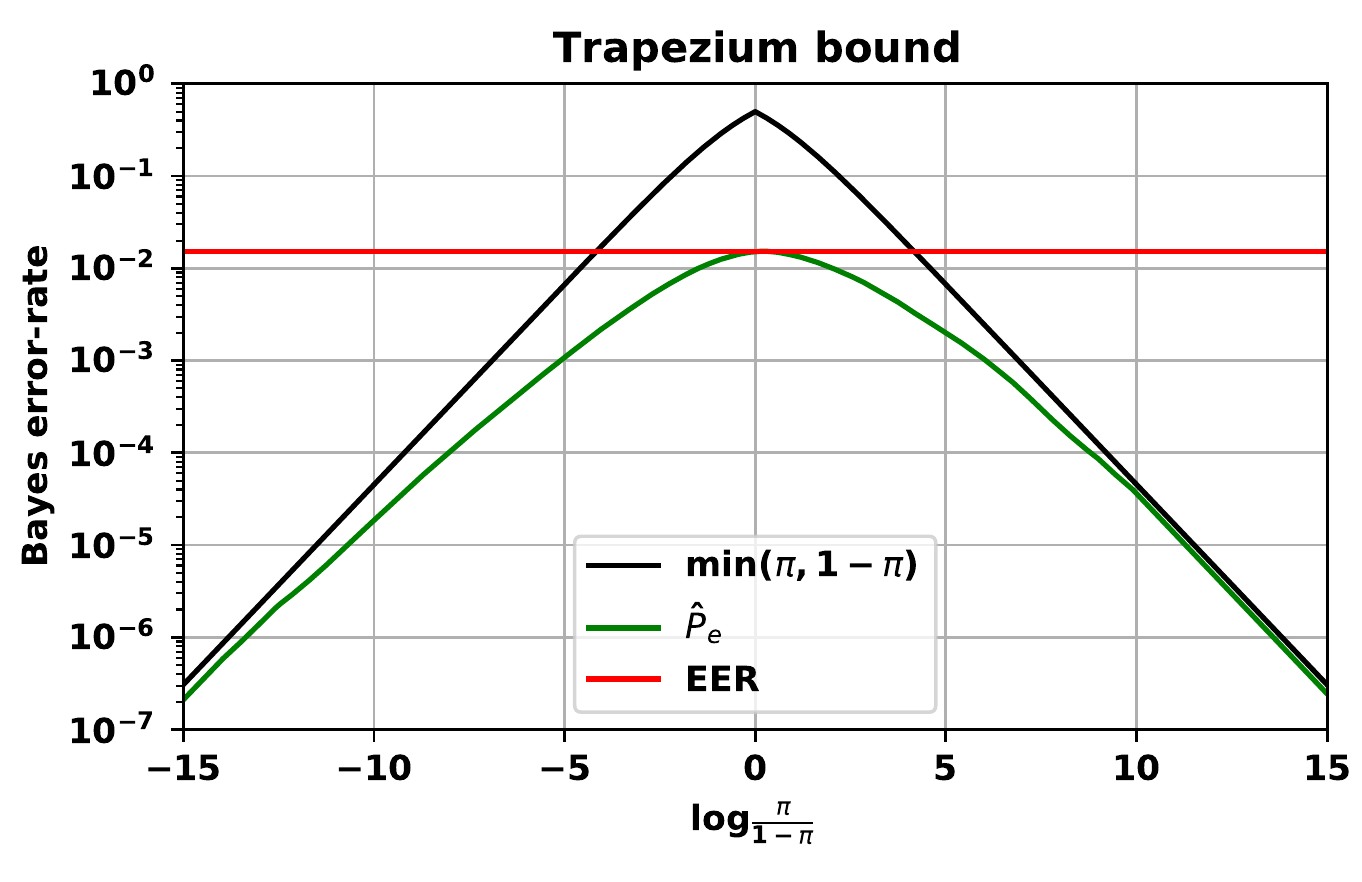}
  \caption{The trapezium bound: $\hat P_e\le\min(\pi,1-\pi,\EER)$. The axes have logit and log scales, to better show small values.}
  \label{fig:trapezium}
\end{figure}
We start with a concise summary of some well-known elements of Bayes decision theory~\cite{DeGroot70}, applied to binary classification and in particular to speaker verification~\cite{bosaris,bookchapter,NikoCSL}. In this section, we analyze Bayes decisions in the idealized case of perfect calibration. In the next section, we generalize to realistic, practical calibration.

Consider a verification trial represented as $(h,s)$, where $h\in\{H_1,H_2\}$ is the unknown hypothesis and $s\in\R$ is the verifier score. Given $s$, we want to make an accept/reject decision. The outcomes $(\acc,H_1)$ and $(\rej,H_2)$ are considered correct, while the two kinds of classification errors are \emph{miss}: $(\rej,H_1)$; and \emph{false-accept}: $(\acc,H_2)$. We now suppose that the decision logic has access to the joint distribution: 
\begin{align}
\label{eq:joint}
P(h,s)&=P(h)P(s\mid h) 
\end{align}
This implies perfect calibration: the prior, $P(h)$, typically supplied by the user, must equal the actual frequency of $H_1$ vs $H_2$ and the likelihoods $P(s\mid H_1)$ and $P(s\mid H_2)$, typically supplied in likelihood-ratio form by the verifier, must equal the actual conditional score distributions. Given these resources, we can compute the hypothesis posterior:
\begin{align}
P(h\mid s) &= \frac{P(s,h)}{P(s,H_1)+P(s,H_2)}
\end{align}     
If we choose to accept or reject, the probability of error is respectively $P(H_2\mid s)$ or $P(H_1\mid s)$. So to minimize the probability of error, make the \emph{Bayes decision}:\footnote{The decision at equiprobable hypotheses can be chosen arbitrarily. }
\begin{align}
\text{decision} &= 
\begin{cases}
\acc & \text{if $P(H_1\mid s) \ge P(H_2\mid s)$} \\
\rej & \text{if $P(H_1\mid s) \le P(H_2\mid s)$}
\end{cases}
\intertext{This can be rewritten, using basic probability rules, as:}
\label{eq:lrdecision}
\text{decision} &= 
\begin{cases}
\acc & \text{if $r \ge \theta_\pi$} \\
\rej & \text{if $r \le \theta_\pi$}
\end{cases}
\end{align}
where $r$ is the \emph{likelihood-ratio} and $\theta_\pi$ is the \emph{Bayes threshold}: 
\begin{align}
\label{eq:lr}
r &= \frac{P(s\mid H_1)}{P(s\mid H_2)} \\ 
\label{eq:BDT}
\theta_\pi &= \frac{P(H_2)}{P(H_1)} = \frac{1-\pi}{\pi}
\end{align}
where $\pi=P(H_1)$ is short-hand. With this optimal decision rule, the probability for a classification error is:\footnote{ The $\hat{}$ on $\hat P_e$ refers to the optimality given by perfect calibration---later, $\tilde P_e$ will be used for the actual error-rate, when perfect calibration cannot be assumed.}
\begin{align}
\label{eq:Pe1}
\begin{split}
\hat P_e(\pi) &= P(\rej,H_1) + P(\acc,H_2) \\
&= \pi P(\rej\mid H_1) + (1-\pi)P(\acc\mid H_2) \\
&= \pi \Pmiss(\theta_\pi) + (1-\pi)\Pfa(\theta_\pi) 
\end{split}
\end{align}
The above conditional error-rates can be expanded as:
\begin{align}
\label{eq:pmiss}
\Pmiss(\theta) &= P(\rej\mid H_1,\theta) = \int_0^{\theta} P(r\mid H_1)\,dr \\
\label{eq:pfa}
\Pfa(\theta) &= P(\acc\mid H_2,\theta) = \int_{\theta}^\infty P(r\mid H_2)\,dr 
\end{align} 
The ROC curve can be generated\footnote{The ROC can also be generated from raw scores, $s$, or from $\log(r)$. For monotonic transforms, $s\to r$ all three ROCs are identical.} by sweeping $\pi$ from 0 to 1 and plotting $\Pmiss$ versus $\Pfa$. As $\pi$ is increased, $\Pmiss$ decreases monotonically and $\Pfa$ increases monotonically.

The full function, $\hat P_e(\pi)$ provides a good answer to the question of optimal verifier accuracy, but let's continue to seek a more concise summary via further analysis. If there is no prior uncertainty the error-rate vanishes: $\hat P_e(0) = \hat P_e(1)= 0$. To understand what happens for $0<\pi<1$, we differentiate $\hat P_e$ w.r.t.\ $\pi$:
\begin{align}
\label{eq:dPedP1}
\hat P_e' &= \Pmiss + \pi\Pmiss' -\Pfa + (1-\pi)\Pfa'
\intertext{where, using~\eqref{eq:BDT} in~\eqref{eq:pmiss} and~\eqref{eq:pfa}:}
\Pmiss' &= \frac{d\Pmiss}{d\theta} \frac{d\theta}{d\pi}
= -\frac{P(\theta\mid H_1)}{\pi^2} \\
\Pfa' &= \frac{d\Pfa}{d\theta} \frac{d\theta}{d\pi}
= \frac{P(\theta\mid H_2)}{\pi^2}
\end{align}
To complete our derivation, we invoke the \emph{calibration property}~\cite{VanLeeuwen_2013} of the likelihood-ratio~\eqref{eq:lr}:
\begin{align}
\label{eq:calprop}
\frac{P(s \mid H_1)}{P(s \mid H_2)} &= r = \frac{P(r \mid H_1)}{P(r \mid H_2)} 
\end{align}
Using this at $r=\theta=\frac{1-\pi}{\pi}$, the 2nd and 4th terms in~\eqref{eq:dPedP1} cancel, to give:
\begin{align}
\hat P'_e(\pi) &= \Pmiss(\theta_\pi) - \Pfa(\theta_\pi)
\end{align} 
After checking that $\hat P''<0$, we see that $\hat P_e$ is concave, with a unique maximum of $\EER=\Pmiss(\theta_{\pi^*}\!)=\Pfa(\theta_{\pi^*}\!)$, where the derivative vanishes at $\pi=\pi^*$:
\begin{align}
\label{eq:maxPe}
\begin{split}
\max_{\pi} \hat P_e &= \pi^*\Pmiss(\theta_{\pi^*}\!) + (1-\pi^*)\Pfa(\theta_{\pi^*}\!)\\ 
&= \pi^*\EER + (1-\pi^*)\EER = \EER 
\end{split}
\end{align}   
This upper-bound property of the EER is not well known. It was mentioned in~\cite{Bernardo94} and received detailed treatment in~\cite{PhD}, but the derivation above is new and we hope more intuitive. Next, we invoke another upper bound~\cite{PhD}:
\begin{align}
\hat P_e(\pi) &\le \min(\pi,1-\pi)
\end{align} 
where the RHS is the probability for error when making a Bayes decision using the prior alone, while the LHS has the advantage of the extra information supplied by the score. The RHS becomes small whenever $\pi\approx0$, or $\pi\approx1$, which shows that the problem becomes easier when the prior uncertainty is low. \\

\noindent In summary of the above results, we can now provide a first answer to our question:
\begin{itemize}[leftmargin=1.5em]
\item[\textbf{Q:}] Out of a hundred trials, how many errors does your speaker verifier make?
\item[\textbf{A:}] The error-rate depends not only on the accuracy of the verifier, but also on the user-supplied prior $\pi$. For perfect calibration, the error-rate, $\hat P_e$, obeys the \emph{trapezium bound}: 
\begin{align}
\label{eq:trapezium}
\hat P_e(\pi) \le \min(\pi,1-\pi,\EER )
\end{align}
wherein the verifier accuracy is represented by the EER, and the user contribution by $\min(\pi,1-\pi)$. See figure~\ref{fig:trapezium} for an example.
\item[\textbf{Q:}] What if calibration is not perfect? 
\item[\textbf{A:}] Calibration can be measured (see the next section), using the same resources as for the ROC, namely a supervised database of scores. Since calibration can be measured, it can be optimized, so that  the actual error-rates can get close to the optimal $\hat P_e$.   
\end{itemize}

\section{Actual Bayes decisions}
\label{sec:aer}
In a real, practical application of a speaker verifier, perfect calibration cannot be assumed, neither for the user-supplied prior, nor for the system-supplied likelihood-ratio. If either of these resources are not perfectly calibrated then the optimal error-rate, $\hat P_e$ of~\eqref{eq:Pe1} becomes a lower bound to the actual error-rate. For bad calibration, the error-rate could become arbitrarily bad, up to a maximum of 1. To see this, consider for example a user that supplies $\pi=1$, when in actual fact only $H_2$ trials arrive at the verifier, in which case every decision will be an error. 

We shall however dismiss the problem of imperfectly calibrated $\pi$, because in this paper, we are interested in quantifying the verifier accuracy and the verifier should not be blamed for user errors. Keep in mind that $\pi$ cannot be extracted by the verifier from the speech inputs, nor can it be learnt from a typical verifier training database, because $\pi$ is entirely dependent on the application. In what follows, we shall assume perfect $\pi$ calibration.

Now, let $P(s\mid h)$ denote the actual conditional score distributions and let $r=\frac{P(s\mid H_1)}{P(s\mid H_2)}$ represent the perfectly calibrated likelihood-ratio, which is in practice \emph{not available} to the decision logic. Instead, we assume that the decision logic has access to a \emph{calibration function}, $\tilde r = f(s)$, that is designed to give $\tilde r \approx r$. (For some systems, the score $s\approx r$ is already well-calibrated, e.g.~\cite{STBU,DCA_PLDA}. For such systems we take $f$ to be the identity function.) Other examples of calibration can be found in~\cite{NikoCSL,bosaris,cal,bayescal,usc,sandro19,sandro19b,sandro20,Ferrer:aslp18}, and in many other systems submitted to speaker recognition evaluations.

To analyze the actual error-rate that results from using $\tilde r$, we use it to replace $r$ in the decision rule~\eqref{eq:lrdecision}, while $\theta_\pi$ is still the same theoretically optimal threshold~\eqref{eq:BDT}. The resulting \emph{actual error-rate} is:\footnote{To analyze prior miscalibration: use the true hypothesis frequencies here, but a miscalibrated prior, $\tilde\pi$ to compute $\theta_{\tilde\pi}$.}
\begin{align}
\label{eq:actPe1}
\tilde P_e(\pi) &= \pi \tildePmiss(\theta_\pi) + (1-\pi)\tildePfa(\theta_\pi) 
\end{align}
where 
\begin{align}
\tildePmiss(\theta) &= \int_0^\theta P(\tilde r\mid H_1)\,d\tilde r \\
\tildePfa(\theta) &= \int_\theta^\infty P(\tilde r\mid H_2)\,d\tilde r 
\end{align} 
To evaluate $\tilde P_e$ in practice, the usual recipe requires the same resource as for the ROC: a supervised database of scores, where $\tilde r$ and $h$ are available for every trial. The conditionals, $P(\tilde r\mid h)$ are now in empirical form (impulses at the data points) and the integrals reduce to counting the errors that result when thresholding all calibrated scores against a set of thresholds of interest. This can be done\footnote{See implementations: \url{github.com/bsxfan/PYLLR}} efficiently by jointly sorting scores and thresholds and retrieving error-rates from the ranks of the thresholds~\cite{bosaris}.

Since generally $\tilde r\ne r$, neither the calibration property~\eqref{eq:calprop}, nor the trapezium bound~\eqref{eq:trapezium} apply to $\tilde P_e$. Since $\hat P_e$ is optimal and $\tilde P_e$ depends on a (hopefully only slightly) suboptimal decision rule, we know that:
\begin{align}
\tilde P_e(\pi) \ge \hat P_e(\pi)
\end{align}
If the calibration is good however, then $\tilde P_e \approx \hat P_e$ and the trapezium will still function as \emph{approximate} upper bound to the actual error-rate: 
\begin{align}
\tilde P_e(\pi) &\approx \hat P_e(\pi) \le \min(\pi,1-\pi,\EER)
\end{align}
For bad calibration however, it is possible for $\tilde P_e$ to deteriorate up to $1$ for some values of $\pi$. See figures~\ref{fig:plda} and~\ref{fig:dca_plda}, in the section on experiments below, where we show some examples of both good and bad calibration. We conclude with further Q\&A: 
\begin{itemize}[leftmargin=1.5em]
    \item[\textbf{Q:}] Which resources are required for good score calibration?
    \item[\textbf{A:}] The same as for any other machine learning problem: algorithms and data---see the references in this section. Methods vary in their data requirements, from large to small and from supervised to unsupervised.
    \item[\textbf{Q:}] It seems that not only score calibration, but also the user's $\pi$ calibration are too hard. Is it not safer and better to directly threshold the scores, with a threshold set at a fixed false-accept rate?
    \item[\textbf{A:}] Easier, maybe. Safer and better, no. Read on.
\end{itemize}

\section{Generalization to expected cost}
Error-rate can be generalized to risk (expected cost), by assigning the respective costs, $\Cmiss,\Cfa>0$ to miss and false-accept errors. The Bayes decision threshold generalizes to:
\begin{align}
\label{eq:thetac}
\theta_\pi^c &= \frac{\Cfa(1-\pi)}{\Cmiss \pi}
\end{align}
For perfect calibration, $\hat P_e$ generalizes to \emph{optimal expected cost}:
\begin{align}
\hat C_e(\pi) &= \Cmiss \pi\Pmiss(\theta_\pi^c) + \Cfa(1-\pi)\Pfa(\theta_\pi^c) 
\end{align}
The trapezium bound becomes~\cite{bosaris}:
\begin{align}
\hat C_e(\pi) \le \min(\Cmiss \pi, \Cfa(1-\pi), R^*)
\end{align} 
where $R^*$ is the \emph{equal risk}, obtained at a threshold, $\theta^*$, such that:
\begin{align}
R^* &= \Cmiss\Pmiss(\theta^*) = \Cfa\Pfa(\theta^*)
\end{align}
In this sense, depending on the ratio $\frac{\Cmiss}{\Cfa}$, any point on the ROC can act as upper bound to the optimal expected cost. For practical calibration, actual expected cost, $\tilde C_e$, can be evaluated in a straight-forward generalization of section~\ref{sec:aer}, see~\cite{bosaris}.

\section{Direct score thresholding}
Given a verifier score, $s$, that is not specifically calibrated to function as likelihood-ratio, consider the decision rule:
\begin{align}
\text{decision} &= 
\begin{cases}
\acc & \text{if $s \ge \bar\theta$} \\
\rej & \text{if $s \le \bar\theta$}
\end{cases}
\end{align}
One way to proceed, when given a \emph{calibration database} of supervised scores, is to choose the score threshold, $\bar\theta=\bar\theta_\pi^c$, to be approximately equivalent to the optimal likelihood-ratio threshold, $\theta_\pi^c$, by doing:\footnote{$\Pmiss,\tildePmiss,\barPmiss$ are computed from respectively $r,\tilde r, s$.}
\begin{align}
\label{eq:mindcf}
\bar\theta_\pi^c &= \argmin_{\theta'} \pi\Cmiss\barPmiss(\theta') + (1-\pi)\Cfa\barPfa(\theta') 
\end{align}
where $\barPmiss,\barPfa$ are defined empirically by counting errors when thresholding scores against $\theta'$ in the calibration database. If we assume $\frac{P(s\mid H_1)}{P(s\mid H_2)}$ is monotonic rising, then the optimal expected cost can be computed even from  \emph{uncalibrated} scores:
\begin{align}
\hat C_e(\pi) &= \min_{\theta'} \pi\Cmiss\barPmiss(\theta') + (1-\pi)\Cfa\barPfa(\theta') 
\end{align}
This is equivalent to defining a calibration function, $\tilde r = f(s)$, non-parametrically, via the PAV algorithm~\cite{cal}. Given  an independent \emph{test database}, the actual risk at $\bar\theta=\bar\theta_\pi^c$:
\begin{align}
\tilde C_e &= \pi\Cmiss\barPmiss(\bar\theta_\pi^c) + (1-\pi)\Cfa\barPfa(\bar\theta_\pi^c) 
\end{align}
gives a more realistic representation of the accuracy at this operating point. Note $\hat P_e$ and $\tilde P_e$ can be obtained from $\hat C_e$ and $\tilde C_e$ by setting $\Cmiss=\Cfa=1$.

Another popular solution is to make use of a calibration database containing only $H_2$ scores (which are often easier to collect than $H_1$ scores) and to choose $\bar\theta$ so that some fixed proportion of $H_2$ scores, say $1\%$ of them, are above the threshold, thus fixing (at least on this data) the false-accept rate. If $H_1$ trials are also available, the miss rate at this threshold can be reported as a representation of accuracy.  

\emph{Does fixing the false-accept rate thus, make this a safe strategy?} No, the rate is fixed exactly \emph{only} for the dataset that was used to choose the threshold. There is no guarantee on unseen data---neither for this method, nor for any other calibration method. Moreover, this method does not lend itself to reporting accuracy on an independent test database, because on the test data, \emph{both} the miss and false-accept rates could change. 

\emph{Surely, this strategy has the advantage that the user does not have to be troubled to provide prior and costs, especially if those could be badly calibrated?} No, ignoring prior and costs, does not cause their essential role in making optimal decisions go away~\cite{Bernardo94,DeGroot70,PTLOS}. Fixing the false-accept rate can be seen as crude way to take the prior and costs into account. Put simply: a bad choice of false-accept rate can have exactly the same effect as a bad choice of $\pi$, $\Cmiss$ or $\Cfa$. 

In summary, we find no advantages to direct score thresholding, other than its perceived simplicity. At best, it is less flexible than likelihood-ratio calibration, because---unlike the adjustable Bayes threshold---the direct score threshold is fixed. Moreover, the fixed-false-accept method has additional disadvantages: it cannot be tested on independent test data; it obfuscates the role of cost and prior; and in general it does something different from Bayes decisions, so that it does \emph{not} minimize error-rates, nor expected costs.

\section{Experimental demonstration}
\begin{figure}[htb!]
  \centering
  \includegraphics[trim = 0 10 0 10, clip, width=\linewidth]{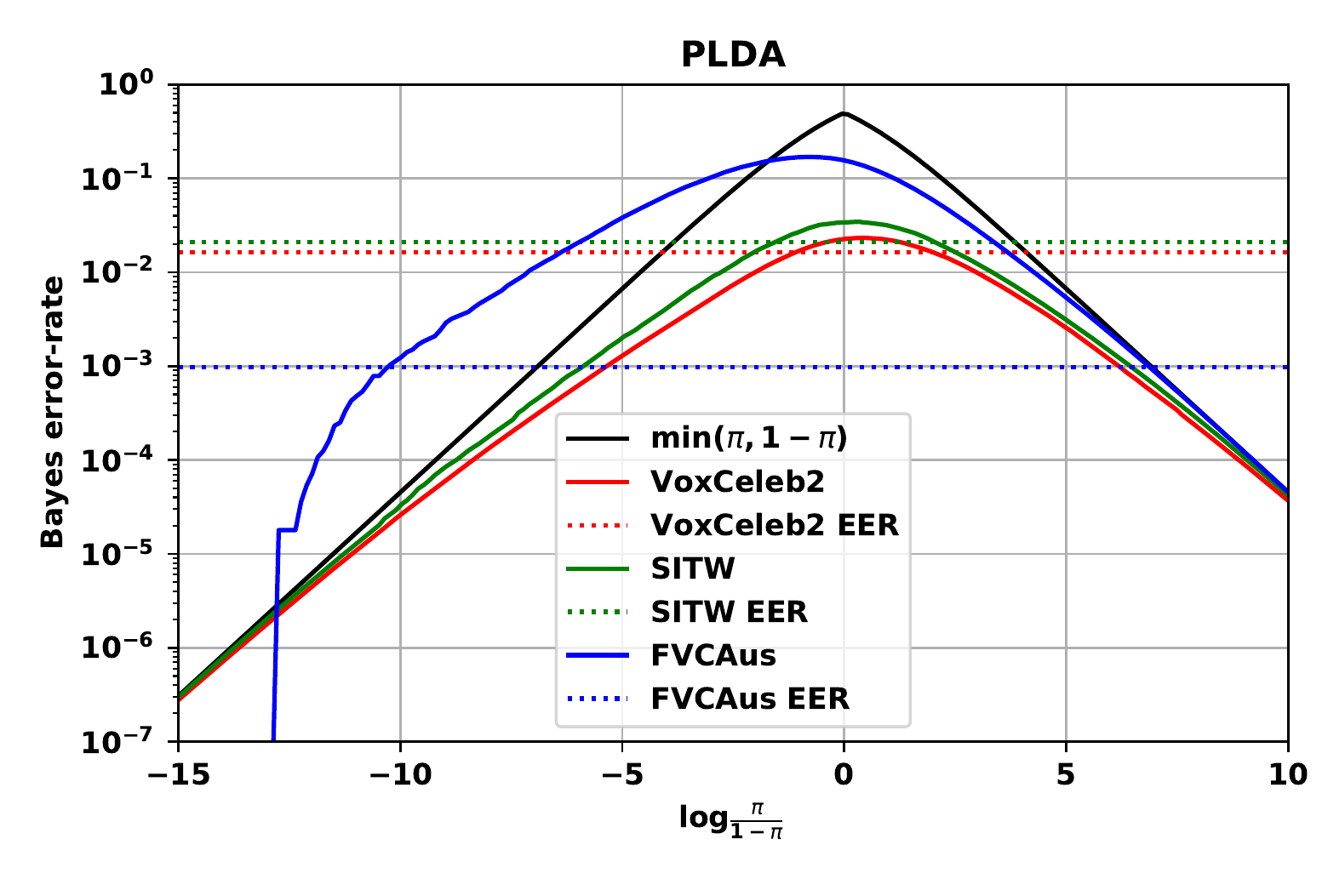}
  \caption{PLDA: $\tilde P_e$ evaluated on three data sets, vs respective trapezium bounds, $\min(\pi,1-\pi,\EER)$.}
  \label{fig:plda}
\end{figure}

Figure~\ref{fig:plda} shows plots\footnote{Code is provided at \url{github.com/bsxfan/PYLLR}.} of the actual Bayes error-rate $\tilde P_e$, for an x-vector speaker verifier, with a standard (generative) PLDA scoring backend, followed by an affine log-likelihood-ratio calibration transform, as implemented in~\cite{DCA_PLDA}. The system was evaluated on three different data sets. On the red and green data sets, the ideal EER bound on $\hat P_e$ is somewhat exceeded, but the prior bound, $\min(\pi,1-\pi)$ is not exceeded. On the blue data set, calibration is very bad, it far exceeds both the EER and the prior bound. The blue data is very clean speech that gives a very low EER (about 0.1\%), but this dataset shift effect causes the calibration to fail. 

\begin{figure}[htb!]
  \centering
  \includegraphics[trim = 0 10 0 10, clip, width=\linewidth]{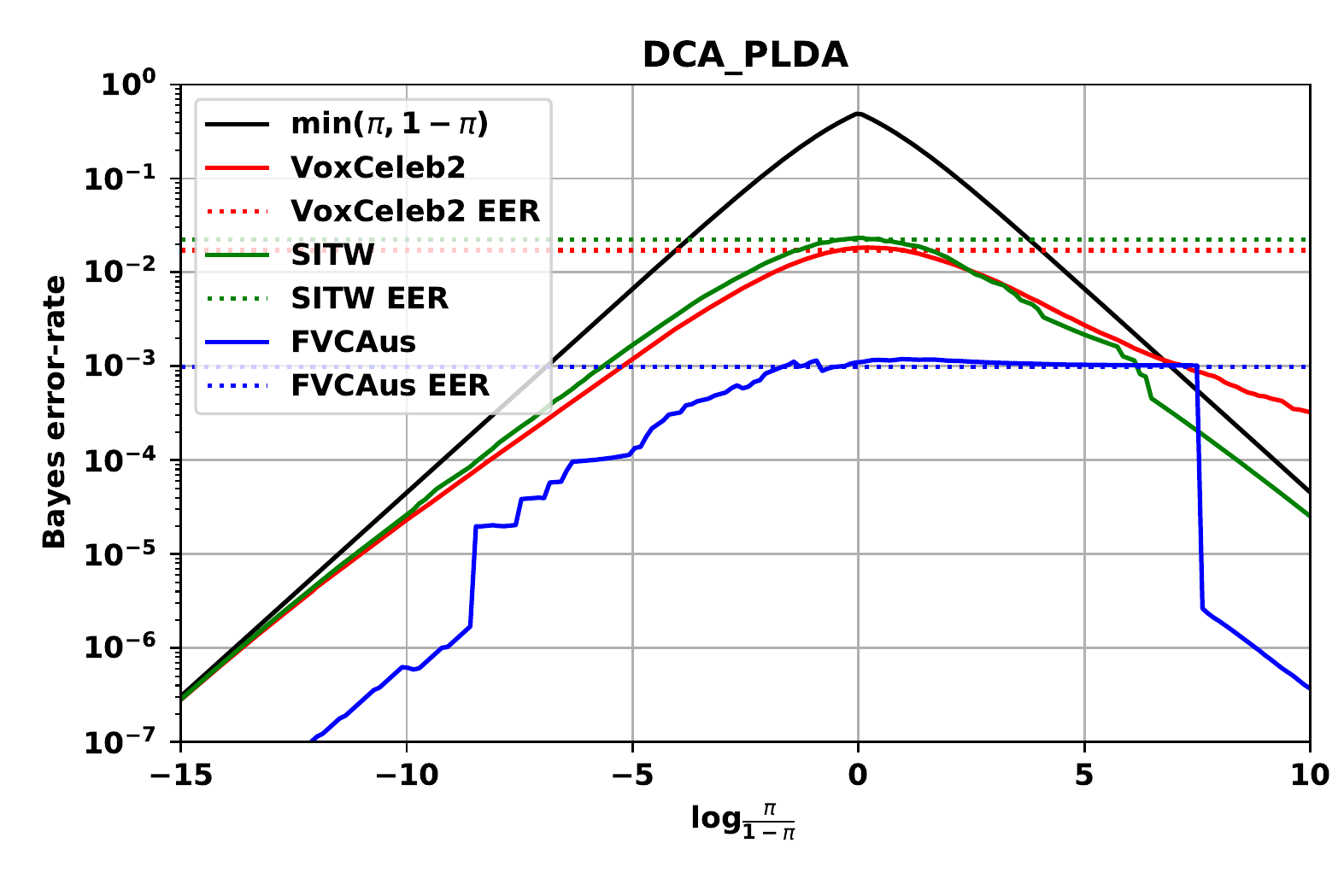}
  \caption{DCA-PLDA: $\tilde P_e$ evaluated on three data sets, vs respective trapezium bounds, $\min(\pi,1-\pi,\EER)$.}
  \label{fig:dca_plda}
\end{figure}

Figure~\ref{fig:dca_plda} shows a \emph{more complex} backend, DCA-PLDA~\cite{DCA_PLDA}, that was discriminatively trained on a \emph{more diverse} database. It shows much better calibration on all three data sets.\footnote{We have preliminary evidence to suggest the violation of the prior bound on the far right may be due to labelling errors in the VoxCeleb 2 test set.}


\section{Conclusion}
Calibrated scores are more versatile than uncalibrated ones. Bayes decisions minimize error-rates and risk. Ad-hoc decisions don't. The Bayes error-rate and risk are more informative to prospective users than $\Pmiss$ and $\Pfa$. Calibration and computing Bayes risk both require some hard work, but the benefits that can be thus obtained can be previewed before investing in this work: the \emph{trapezium upper bound} on the Bayes error-rate (risk) can be computed from the EER (ROC) and these can be computed from uncalibrated scores with standard methods. Although this bound applies only to perfectly calibrated scores, it provides an ideal that can be pursued by practical calibration strategies. 

\section{Acknowledgements}
We would like to thank Corne van Biljon of Gendac (Pty) Ltd, for insisting on a simple and direct answer to his question, ``Out of a hundred trials, how many errors would a typical speaker verifier make?''

\bibliographystyle{IEEEtran}

\bibliography{eerbib}

\end{document}